\documentclass[runningheads]{llncs}

\usepackage[acronym, shortcuts]{glossaries-extra}

\usepackage{graphicx}
\usepackage[hidelinks]{hyperref}
\usepackage{multirow}
\usepackage[moderate, tracking=normal, leading=normal]{savetrees}
\usepackage{subcaption}
\usepackage{soul}
\usepackage{xcolor}

\usepackage[strict]{changepage}

\usepackage{framed}

\definecolor{formalshade}{rgb}{0.85,1,0.85}
\definecolor{darkblue}{rgb}{0.0,0.6,0.30}
\newenvironment{formal}{%
  \MakeFramed{\advance\hsize-\width\FrameRestore}%
  \noindent\hspace{-4.55pt}
  \begin{adjustwidth}{}{7pt}%
}
{%
  \end{adjustwidth}\endMakeFramed%
}

\usepackage{tikz}
\newcommand*\circled[1]{\tikz[baseline=(char.base)]{
            \node[shape=circle,draw,inner sep=0.5pt] (char) {\scriptsize #1};}}

\glssetcategoryattribute{acronym}{nohyper}{true}
\setabbreviationstyle[acronym]{long-short}

\newacronym{ai}{AI}{Artificial Intelligence}
\newacronym{ann}{ANN}{Artificial Neural Networks}
\newacronym{asr}{ASR}{Attack Success Rate}
\newacronym{ads}{ADS}{Anomaly Detection System}
\newacronym{can}{CAN}{Controller Area Network}
\newacronym{cnn}{CNN}{Convolutional Neural Network}
\newacronym{cpu}{CPU}{Central Processing Unit}
\newacronym{crc}{CRC}{Cyclic Redundancy Check}
\newacronym{ddos}{DDoS}{Distributed Denial of Service}
\newacronym{dnn}{DNN}{Deep Neural Network}
\newacronym{ecu}{ECU}{Electronic Control Unit}
\newacronym{ev}{EV}{Electric Vehicle}
\newacronym{far}{FAR}{False Acceptance Rate}
\newacronym{fgsm}{FGSM}{Fast Gradient Sign Method}
\newacronym{fn}{FN}{False Negative}
\newacronym{fp}{FP}{False Positive}
\newacronym{gan}{GAN}{Generative Adversarial Network}
\newacronym{gps}{GPS}{Global Positioning System}
\newacronym{gpu}{GPU}{Graphics Processing Unit}
\newacronym{gru}{GRU}{Gated Recurrent Unit}
\newacronym{ids}{IDS}{Intrusion Detection System}
\newacronym{imu}{IMU}{Inertial Measurement Unit}
\newacronym{lstm}{LSTM}{Long Short-Term Memory}
\newacronym{lstmfcn}{LSTM-FCN}{Long Short-Term Memory Fully Convolutional Network}
\newacronym{ml}{ML}{Machine Learning}
\newacronym{mlp}{MLP}{Multi-Layer Perceptron}
\newacronym{dl}{DL}{Deep Learning}
\newacronym{obd}{OBD}{On-Boad Diagnostic}
\newacronym{oem}{OEM}{Original Equipment Manufacturer}
\newacronym{rnn}{RNN}{Recurrent Neural Network}
\newacronym{shap}{SHAP}{SHapley Additive exPlanations}
\newacronym{svdd}{SVDD}{Support Vector Domain Description}
\newacronym{svm}{SVM}{Support Vector Machine}
\newacronym{tn}{TN}{True Negative}
\newacronym{tp}{TP}{True Positive}
\newacronym{tpm}{TPM}{Trusted Platform Module}
\newacronym{ubm}{UBM}{Universal Background Model}
\newacronym{ubi}{UBI}{Usage Based Insurance}
\newacronym{uds}{UDS}{Unified Diagnostic Services}
\newacronym{mitm}{MitM}{Man-in-the-Middle}
\newacronym{rpm}{RPM}{Revolutions Per Minute}
\newacronym{rf}{RF}{Random Forest}
\newacronym{gmm}{GMM}{Gaussian Mixture Model}
\newacronym{knn}{kNN}{k-Nearest Neighbors}
\newacronym{xai}{XAI}{Explainable Artificial Intelligence}

\usepackage{pifont}
\newcommand{\cmark}{\ding{51}}%
\newcommand{\xmark}{\ding{55}}%

\begin{document}

\title{FaultGuard: A Generative Approach to Resilient Fault Prediction in Smart Electrical Grids}
\titlerunning{FaultGuard}

\author{Emad~Efatinasab\inst{1} \and
Francesco~Marchiori\inst{2} \and
Alessandro~Brighente\inst{2} \and
Mirco~Rampazzo\inst{1} \and
Mauro~Conti\inst{2, 3}}
\authorrunning{E. Efatinasab et al.}
\institute{University of Padova, Department of Information Engineering \and
University of Padova, Department of Mathematics \and
Delft~University~of~Technology, Faculty of Electrical Engineering, Mathematics and Computer Science \\
\email{\{emad.efatinasab, francesco.marchiori.4\}@phd.unipd.it}
\email{\{alessandro.brighente, mirco.rampazzo, mauro.conti\}@unipd.it}}
\maketitle             
\begin{abstract}
Predicting and classifying faults in electricity networks is crucial for uninterrupted provision and keeping maintenance costs at a minimum.
Thanks to the advancements in the field provided by the smart grid, several data-driven approaches have been proposed in the literature to tackle fault prediction tasks.
Implementing these systems brought several improvements, such as optimal energy consumption and quick restoration.
Thus, they have become an essential component of the smart grid.
However, the robustness and security of these systems against adversarial attacks have not yet been extensively investigated.
These attacks can impair the whole grid and cause additional damage to the infrastructure, deceiving fault detection systems and disrupting restoration.

In this paper, we present \textbf{FaultGuard}, the first framework for fault type and zone classification resilient to adversarial attacks.
To ensure the security of our system, we employ an Anomaly Detection System (ADS) leveraging a novel Generative Adversarial Network training layer to identify attacks.
Furthermore, we propose a low-complexity fault prediction model and an online adversarial training technique to enhance robustness.
We comprehensively evaluate the framework's performance against various adversarial attacks using the IEEE13-AdvAttack dataset, which constitutes the state-of-the-art for resilient fault prediction benchmarking.
Our model outclasses the state-of-the-art even without considering adversaries, with an accuracy of up to 0.958.
Furthermore, our ADS shows attack detection capabilities with an accuracy of up to 1.000.
Finally, we demonstrate how our novel training layers drastically increase performances across the whole framework, with a mean increase of 154\% in ADS accuracy and 118\% in model accuracy.
\end{abstract}

\section{Introduction}
\label{sec:introduction}

Smart grids represent a transformative paradigm in the realm of energy distribution~\cite{bayindir2016smart,muqeet2023state}.
Through advanced technologies, they aim to enhance electrical grids' efficiency, reliability, and sustainability.
Unlike traditional power distribution systems, smart grids leverage real-time data, communication networks, and intelligence control mechanisms to optimize electricity generation, distribution, and consumption.
As such, they enable a bidirectional flow of information between utilities and customers, forming a responsive energy ecosystem~\cite{muqeet2023state}.
The significance of smart grids lies in their ability to address the challenges posed by the evolving energy landscape.
Indeed, they facilitate the integration of renewable resources such as solar and wind, mitigating the impact of their variability and contributing to the overall sustainability of the energy sector.
Given their importance in the current energy landscape, ensuring the security of smart grids is imperative.
Indeed, their interconnection and dependence on digital communication expose them to potential cyber threats and vulnerabilities~\cite{nafees2023smart}.
As smart grids increasingly rely on data-driven technologies, robust security measures are indispensable to safeguard confidentiality, integrity, and availability across the energy infrastructure.
Despite the many papers in the literature proposing new models and methodologies for various aspects of smart grids~\cite{inproceedings,https://doi.org/10.1049/hve.2016.0005,en16052280,en13092149,https://doi.org/10.1049/iet-gtd.2016.0364,inproceedings2,en13133460,STEFANIDOUVOZIKI2022108031,7398152}, a notable gap exists in addressing their security considerations.
As the proposed models increasingly rely on \ac{ai} and \ac{ml}, it is imperative to address the inherent vulnerabilities that these methodologies suffer from, such as adversarial attacks.

\paragraph{Contribution.}
To reduce this gap in the literature, we propose \textbf{FaultGuard}, a framework for fault type and fault zone prediction in smart grids resilient to adversarial attacks.
Unlike many studies focusing on enhancing predictive capabilities, we emphasize resiliency and incorporate robust security layers.
In particular, compared to the state-of-the-art~\cite{mainpaper}, we \textit{(i)} add an \ac{ads} and \textit{(ii)} employ adversarial training in different parts of the system.
The \ac{ads} detects adversarial attacks toward the fault prediction system, showing an accuracy of up to 1.000 when paired with an adversarial learning training technique.
Our new learning technique shows an average improvement of the \ac{ads} of 154\% compared to its counterpart that has not been trained with adversarial learning.
We then develop a low-complexity fault prediction system outperforming the state-of-the-art~\cite {mainpaper}.
To increase the resilience of our fault prediction system against adversarial attacks, we propose and employ \textit{online adversarial training} during its training phase.
This procedure shows a mean increase in the model's accuracy of up to 118\% when the system is under attack.
We evaluate our model on the IEEE13-AdvAttacks dataset~\cite{mainpaper}, a simulated dataset based on the IEEE-13 test node feeder.
Our results show that our model outclasses the state-of-the-art, reaching an accuracy of up to 0.958.
Our contributions can be summarized as follows.

\begin{itemize}
    \item We propose \textbf{FaultGuard}, a resilient framework for predicting fault types and zones in smart grids capable of withstanding adversarial attacks.
    \item We propose a single-layer \ac{gru} architecture that outclasses the state-of-the-art in fault type and fault zone prediction.
    \item We propose an \ac{ads} capable of detecting complex adversarial attacks generated with different amounts of adversarial noise.
    \item We propose an online adversarial training technique, showing how including a subset of adversarial samples in the training process drastically increases the accuracy of the models under attack.
    \item We evaluate our models, attacks, and defenses on a publicly available dataset, showing the efficacy of the attacks in unrestricted scenarios and the capabilities of our defenses.
    \item We make the code of our systems, attacks, and the dataset available at: \url{https://anonymous.4open.science/r/FaultGuard-1518}.
\end{itemize}

\paragraph{Organization.}
The paper is organized as follows.
In Section~\ref{sec:relatedworks}, we mention the challenges and limitations of the related works in the literature.
Our system and threat models are proposed in Section~\ref{sec:systhreat}.
The methodology for our attacks is discussed in Section~\ref{sec:attacks}, while the details of our model implementation are discussed in Section~\ref{sec:delamain}.
In Section~\ref{sec:evaluation}, we evaluate our model, attacks, and defenses.
We report the takeaways of this study in Section~\ref{sec:takeaways}, and Section~\ref{sec:conclusions} concludes this work.
\section{Related Works}
\label{sec:relatedworks}

While the literature has extensively discussed and implemented fault prediction models on smart grids, their security and robustness have not been thoroughly studied.
Indeed, these models have been shown to be vulnerable to adversarial attacks.
For instance, Ardito et al.~\cite{mainpaper} investigated the robustness of fault type and zone classification systems against adversarial attacks.
They conducted evaluations through dataset releases, benchmarking, and assessments of smart grid failure prediction systems under adversarial assaults.

Numerous papers have delved into fault detection and classification methodologies within Smart Grids~\cite{inproceedings,https://doi.org/10.1049/hve.2016.0005,en16052280,en13092149,https://doi.org/10.1049/iet-gtd.2016.0364,inproceedings2,en13133460,STEFANIDOUVOZIKI2022108031,7398152}.
As outlined by Saha et al.~\cite{saha2009fault}, the categorization of fault location methodologies in power systems includes traditional, observant, and intelligent approaches.
This paper specifically focuses on intelligent approaches for fault detection, utilizing smart sensors or expert systems.
These intelligent methods involve various techniques, such as expert systems, \ac{ml}, and \ac{dl}, all aimed at identifying faults within the system.
Indeed, \acp{ann} have been extensively explored in the literature for identifying and predicting faults~\cite{al2003fault,aslan2012alternative,4441599,iet:/content/journals/10.1049/oap-cired.2017.0007,FARIAS201820,5212044,905590,8586471}.
Shadi et al.~\cite{SHADI2022107399} leveraged \ac{rnn} and \ac{lstm} models within a real-time hierarchical architecture to accurately pinpoint and localize faults.
Bhattacharya et al.~\cite{8372054} developed a framework for intelligent fault analysis, leveraging \acp{svm} and \acp{lstm}.
Zhang et al.~\cite{9072421} introduced a method leveraging the attention mechanism, Bidirectional \ac{gru}, and a dual structure network to analyze data from diverse perspectives.
Thukaram et al.~\cite{1413307} proposed a hybrid approach combining \ac{svm} and \ac{ann} architectures.
In their method, the \ac{svm} streamlines the relationship between measurements and fault distance. 
Tree-based methods like \acp{rf} have emerged as highly favored techniques for fault location due to their versatility and low variance~\cite{okumus2021random,s22020458}.
Also, Sapountzoglou et al.~\cite{SAPOUNTZOGLOU2020106254} proposed a gradient-boosting tree model to detect, identify, and localize faults within low-voltage smart distribution grids. 
Wilches-Bernal et al.~\cite{9817473} introduced an innovative fault location and classification algorithm, leveraging mathematical morphology in conjunction with \acp{rf}.
Majidi et al.~\cite{6915710} introduced a fuzzy-c clustering approach to identify potential fault points. 
Ghaemi et al.~\cite{GHAEMI2022107766} introduced an ensemble approach to enhance the precision of fault node localization.
Their method is designed to leverage the strengths of \acp{svm}, \acp{knn}, and \acp{rf}.
\section{System and Threat Model}
\label{sec:systhreat}

We now delve into the system and threat model of our study.
In the former, we disclose the standard functionality of the system in adversary-free environments.
In the latter, we discuss the possible attacker's capabilities and the assumption of their system knowledge.

\paragraph{System Model.}
In an unthreatened scenario (i.e., without attackers aiming to disrupt the system), the model inputs the data from the smart grid infrastructure.
We assume having two fault prediction models, one for each task: fault type prediction and fault zone prediction.
In the former, the model objective is to determine the type of voltage sags faults, which can be asymmetric phase-to-phase (LL), single-phase-to-ground (LG), two-phase-to-ground (LLG), or symmetric three-phase-to-ground (LLLG or LLL).
In the latter, the model objective is determining the geographical zone where the fault occurred.
We assume the models have been trained on an uncorrupted dataset and are finally deployed into the system.

\paragraph{Threat Model.}
As we aim to provide efficient defenses against adversaries targeting \ac{ml} models in the smart grid, we delineate our threat model encompassing the most favorable scenarios for the attacker.
Thus, while aiming to compromise the fault prediction model, we assume the adversary can successfully infiltrate the system and inject data into the grid.
There are various ways in which an attacker can achieve this, as exploiting known or new vulnerabilities was demonstrated to be an effective way to gain remote access~\cite{chen2011lessons,sullivan2017cyber}.
Once an adversary has gained access to the infrastructure, they aim to compromise fault type or fault zone prediction using adversarial examples.
In the former scenario, the attacker intends to cause the misclassification of potential faults, potentially prompting inappropriate recovery actions by grid operators, leading to catastrophic consequences.
In the latter scenario, the adversary manipulates fault prediction models, targeting fault zone prediction in smart grids.
This results in recovery teams being dispatched erroneously to the wrong zone, amplifying the impact on operational efficiency and necessitating robust security measures to safeguard smart grid applications. 

We can define two scenarios based on the attacker’s knowledge of the exchanged data and the smart grid models.
\begin{itemize}
    \item \textit{White-box Scenario}: the attacker has access to the data used for testing the model and the model's architecture and parameters.
    This is the most favorable scenario for the attacker, who can leverage this intelligence to craft powerful adversarial samples.
    Furthermore, having access to the model weights allows the adversary to tune the attack parameters offline.
    \item \textit{Gray-box Scenario}: the attacker has access to the data used for testing the model, but not the model's architecture or parameters.
    This scenario is more challenging for an adversary aiming to use adversarial samples, as it would require them to be transferable among different model architectures.
    However, several studies have demonstrated the difficulties of this task and showed the inefficiency of using surrogate models for generating adversarial noise to test samples~\cite{alecci2023dumb,apruzzese2023real}.
\end{itemize}
It is worth noting that while the white-box scenario is the most favorable from the attacker's perspective, the gray-box scenario is more achievable in real-world implementations.
Indeed, an adversary can obtain the model architecture from either \textit{(i)} the known implementation disclosed by the manufacturer, or \textit{(ii)} having direct access to the system input/output and use model extraction techniques~\cite{gong2020model,jagielski2020high}.
As such, model parameters can be protected by \textit{(i)} not publicly disclosing the model architecture and training dataset used and \textit{(ii)} using model obfuscation techniques.
Instead, gaining data is a more accessible technique for the attacker, as many entry points are present across the infrastructure.
Indeed, many IoT devices and networks compose the smart grid.
With the integration of data coming from sustainable energy producers, adversaries can collect data in various parts of the system.
\section{Attacks}
\label{sec:attacks}

We now discuss the attacks that we employ against fault type and zone prediction systems in smart grids.
Our attacks are different depending on the assumptions of the attacker's knowledge, namely, white-box scenario (Section~\ref{subsec:white}) and gray-box scenario (Section~\ref{subsec:gray}).

\subsection{White-box Scenario}
\label{subsec:white}

In our white-box threat model, the adversary possesses full knowledge of the data and the trained model.
Thus, we analyze prominent adversarial attacks to reveal vulnerabilities in \ac{ml} models.
We focus on specific attacks highlighted in the literature for their significance and capacity to uncover weaknesses.

\begin{itemize}
  \item \textit{Fast Gradient Sign Method (FGSM):} swiftly crafts adversarial examples by leveraging the sign of the gradient of the loss function. Recognized for computational efficiency, it is a foundational benchmark for evaluating model robustness~\cite{goodfellow2015explaining}.
  \item \textit{Basic Iterative Method (BIM):} extends FGSM through an iterative application, introducing small perturbations at each step to enhance attack potency. Provides insights into cumulative perturbation effects for nuanced robustness evaluation~\cite{kurakin2017adversarial}.
  \item \textit{Carlini \& Wagner (CW):} a sophisticated attack that formulates adversarial example crafting as an optimization problem, seeking minimal perturbations for misclassification with minimal perceptibility. Challenges models with minimal perturbations, assessing resistance against imperceptible adversarial examples~\cite{7958570}.
  \item \textit{Randomized Fast Gradient Sign Method (RFGSM):} introduces randomness into FGSM iterations by incorporating random noise, enhancing attack diversity. Explores the impact of variability in adversarial perturbations, providing insights into model robustness against unpredictable attacks~\cite{tramèr2020ensemble}.
  \item \textit{Projected Gradient Descent (PGD):} employing an iterative optimization approach akin to BIM, PGD includes a projection step to confine perturbations within a defined constraint set. It stands out for crafting potent adversarial examples, allowing rigorous examination of model robustness under stringent conditions~\cite{madry2019deep}.
\end{itemize}

\subsection{Gray-box Scenario}
\label{subsec:gray}

In the gray-box scenario, the adversary has gained access solely to the data and cannot get access to the prediction models.
Exploiting this limited access, the adversary employs a \ac{gan} to synthesize malicious data that closely mirrors authentic instances.
\acp{gan} are a class of artificial intelligence algorithms that consist of two neural networks, a generator and a discriminator, trained simultaneously to generate realistic data.
In this way, the attacker can lead the fault prediction model towards the detection of a specific fault type or zone.
This evasion strategy involves training a \ac{gan} model on real data, enabling synthetic data generation that resembles legitimate smart grid data.
By successfully training the \ac{gan}, the adversary acquires a powerful tool (i.e., the trained generator), which can then be employed to inject the generated data into the smart grid system.
Introducing maliciously generated data designed to mimic real data poses a nuanced challenge, showcasing the adversarial capabilities of \acp{gan} in evading detection.
\section{FaultGuard}
\label{sec:delamain}

\begin{figure}[!htpb]
    \centering
    \includegraphics[width=.8\textwidth]{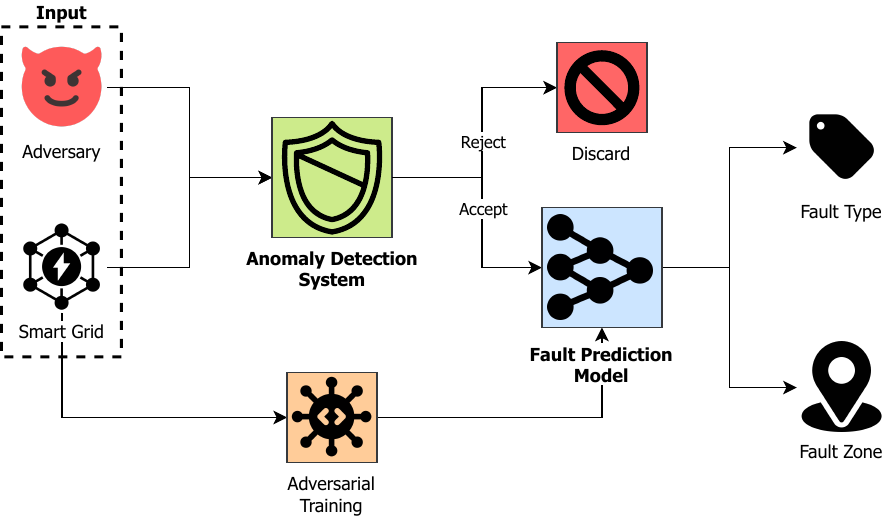}
    \caption{FaultGuard framework.}
    \label{fig:pipeline}
\end{figure}

In this section, we present our proposed FaultGuard framework, which is graphically shown in Fig.~\ref{fig:pipeline}.
We consider two primary data sources for prediction: the legitimate sensor data gathered in the smart grid and the malicious data injected by possible adversaries.
We employ an \ac{ads} for detecting adversarial samples in the input data.
If samples are detected as malicious, they are discarded.
Instead, if the data appears legitimate, they are fed as input to the fault prediction model.
Furthermore, we employ online adversarial training to boost our system resilience towards possible attacks.
Finally, our model generates a prediction for each task it is trained on: fault type and zone prediction.

We present our \ac{gan}-based \ac{ads} in Section~\ref{subsec:ad}, and the fault prediction system in Section~\ref{subsec:Faultgurad}.
While we list the components of our pipeline in order or appearance, it is worth noting that, in real-world scenarios, the first step would be training the fault prediction system.
Indeed, our \ac{ads} uses the trained prediction model in its implementation.
After training both the model and the \ac{ads}, components can be organized as detailed in Fig.~\ref{fig:pipeline}.

\subsection{Anomaly Detection System}
\label{subsec:ad}

As shown in Fig.~\ref{fig:pipeline}, we employ an \ac{ads} before feeding the input to our fault prediction system.
The aim of the \ac{ads} is to detect and reject adversarial attacks while allowing legitimate samples.
We use a \ac{gan} to achieve this.
Our \ac{gan} model is characterized by neural networks featuring linear input and output layers.
In particular, we leverage the discriminator for anomaly detection after training.

\paragraph{Architecture.}
The generator model creates synthetic data that mimic legitimate data patterns.
It achieves this through four fully connected layers, with neurons varying from 51 (i.e., the number of features) to 128.
The discriminator model serves a dual purpose as an \ac{ads} and an authenticity evaluator.
It is tasked with evaluating the genuineness of incoming data by discerning whether it is authentic grid data (real) or artificially generated by the generator model (fake).
Comprising five fully connected layers, the discriminator's neuron count ranges from 51 to 512.
More details on our implementation and hyper-parameters are publicly available in our GitHub repository.

\begin{figure}[!htpb]
    \centering
    \includegraphics[width=.8\textwidth]{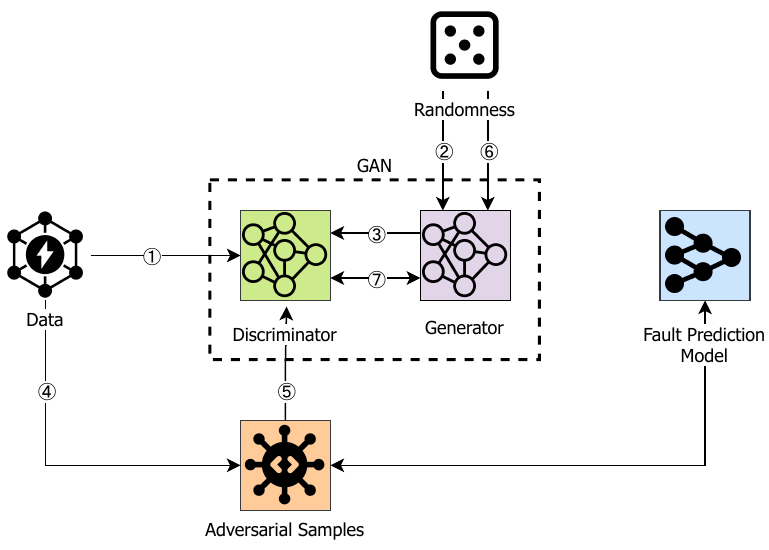}
    \caption{Schema of our \ac{gan}-based \ac{ads} training process for each epoch.}
    \label{fig:training}
\end{figure}

\paragraph{Training.}
Recognizing the imperative to fortify the discriminator's capabilities, we introduce a novel layer of training in the traditional \ac{gan} training process.
An overview of this process is shown in Fig.~\ref{fig:training}.
The first training steps adhere to the standard procedure and, as such, start with training the discriminator on real data (step \circled{1}).
Once the discriminator's loss is backpropagated, we generate fake data with the generator.
We do this by starting with a random tensor of latent inputs, which the generator model consequently processes to create the fake inputs (step \circled{2}).
We evaluate these samples with the discriminator and subsequently backpropagate the loss (step \circled{3}).
Before proceeding with the training of the generator, we first add our novel layer of training.
To increase the adversarial detection capabilities of our discriminator, we feed adversarial samples with the FGSM and BIM attack (step \circled{4}).
These samples are derived from the real data the discriminator was previously trained on and generated through the fault prediction model.
In this way, the discriminator can increase its capabilities in detecting FGSM and BIM samples (step \circled{5}); however, the transferability property of these attacks allows the \ac{ads} also to detect other types of attacks~\cite{alecci2023dumb}.
We call this layer of training \textit{adversarial learning}, whose contributions are evaluated in Section~\ref{subsec:faultguardevaluation}.
Finally, we train the generator by generating another batch of fake data (step \circled{6}), computing the loss on the discriminator, and backpropagating it to the generator model (step \circled{7}).
This process is repeated for 100 epochs, with a learning rate of $2 \times 10^{-4}$ for each model.

\subsection{Fault Prediction System}
\label{subsec:Faultgurad}

In this section, we comprehensively explore the novel model we introduced for fault type and zone prediction tasks.
The fault prediction system constitutes the ultimate stage in our framework, working with data that has undergone thorough legitimacy verification by the \ac{ads}.
While the preceding work, delineated in the dataset introduction paper, included a benchmarking model utilizing a \ac{mlp}, its performance was not the primary focus of the study~\cite{mainpaper}.
Indeed, it is worth noting that the causality between consecutive data samples is lost by using a \ac{mlp}.
The oversight in considering the temporal relationships among data points could contribute to the sub-optimal performance observed in benchmarking the model.

\paragraph{Architecture.}
Recognizing the importance of enhanced performance in real-world scenarios, we propose a novel model tailored for superior efficacy in fault type and fault zone classification.
Focusing on practicality and efficiency, we design our fault prediction model to maintain simplicity and minimize computational demands.
The core architecture of our model centers on a one-layer bidirectional \ac{gru} comprising 220 neurons.
This design enables the model to adeptly capture dependencies in both forward and backward directions within the temporal sequence.
Following the \ac{gru} layer, we introduced a dropout layer with a rate of 0.5.
This layer mitigates overfitting and enhances the model's resilience to the subtle changes that characterize adversarial samples. 
Post-dropout, the output of the \ac{gru} layer transforms a linear layer housing 440 neurons.
Subsequently, an element-wise sigmoid activation function is applied, yielding an output vector with dimensions equal to the number of considered classes.

\paragraph{Training.}
We use cross-entropy as a loss function for our model training.
Our optimization process employs the Adam optimizer with a learning rate of $1 \times 10^{-3}$, a value empirically selected for efficient convergence. 
In our case, the training regimen spans a fixed number of epochs set to 80.
While employing the \ac{ads} contributes a significant layer of security, we have developed an additional strategy to fortify our system.
In fortifying the resilience of our fault prediction model, we have devised a strategy involving adversarial training, specifically incorporating both BIM and FGSM attacks.
Traditionally, adversarial training involves generating attacks on a pre-trained model using diverse algorithms and subsequently augmenting the training dataset.
In contrast, our approach integrates the attack generators directly into the model training process.
We call this approach \textit{online adversarial training}.
In this scenario of online adversarial training, the adversary adapts over time, becoming more sophisticated as the model improves.
This adaptive training strategy challenges the model with increasingly difficult adversarial examples, forcing it to improve its robustness continually. 
Our adversarial training methodology entails a single, comprehensive training loop.
The model is sequentially trained within this loop on real and adversarial data generated by FGSM and BIM attacks.
We dynamically create adversarial inputs for each batch of inputs and corresponding labels using FGSM and BIM attacks with a specific epsilon value of 0.2.
This value controls the amount of noise added to the samples and, in our case, is large enough to cause misclassification but small enough to maintain some level of perceptibility in the perturbed examples.
The model's robustness is tested against realistic perturbations within a reasonable range of what an attacker might apply in real-world scenarios by choosing a moderate epsilon value (i.e., from 0.05 to 0.5).
Subsequently, the model undergoes forward passes using these adversarial inputs, and losses for both attacks are computed against the original labels.
The backward pass is executed following the forward passes, and the model's gradients are updated using the optimizer.
We aggregate the total training loss for the epoch by summing the adversarial losses obtained from FGSM and BIM attacks, scaling the cumulative loss based on the batch size processed during each iteration.
This step ensures an appropriate scaling of the total loss relative to the batch size.
The accumulated loss is the foundation for computing the average training loss after each epoch.
By integrating this new layer of training into the model, we can enhance its performance on real-world data, akin to a form of data augmentation.
\section{Evaluation}
\label{sec:evaluation}

We now delve into the evaluation of the attacks and FaultGuad system.
Our evaluation comprehends all scenarios detailed in the previous sections.
We first give details on the dataset used for our evaluation in Section~\ref{subsec:dataset}.
To provide a baseline evaluation to discuss the success of our attacks and defenses, we evaluate our models on different tasks in Section~\ref{subsec:baselineevaluation}.
We then evaluate our attacks in Section~\ref{subsec:attacksevaluation}, and finally study the capabilities of FaultGuard in Section~\ref{subsec:faultguardevaluation}.

\subsection{Dataset}
\label{subsec:dataset}

In the electrical industry, a variety of simulation programs are being used to address fault prediction challenges, including PSCAD~\cite{8341819,6850055}, MATLAB Simulink~\cite{8627971}, RSCAD~\cite{8304532}, and MATPOWER~\cite{5767534}.
Despite the extensive use of these simulation tools in smart grid failure prediction systems, there is a lack of publicly available datasets generated from these tools.
So we turn to the dataset introduced by Ardito et al.~\cite{mainpaper}, the only publicly available dataset including substantial simulated fault data rooted in the IEEE-13 test node feeder.
The IEEE-13 node test feeder includes a 4.16~kV voltage generator, 13 buses for fault simulation, and three-phase signal measurement.
The distribution system is divided into four zones, which are used to identify the location of a fault that has occurred.
This dataset comprises 51 features and two target classes: fault label and fault zone, incorporating both traditional and renewable energy sources.
It has been carefully curated to serve as a benchmark for assessing the effectiveness of adversarial attacks against fault prediction systems in smart electrical grids.
Moreover, we introduce a robust windowing technique to handle our data effectively.
This involves partitioning the dataset into segments, each of a predefined size.
These windows are created by iteratively traversing the data with a step size equal to half of the window size.
Specifically, we choose a window size of 16 seconds for our dataset.
To enhance the dataset's quality and optimize it for our prediction models, we conduct essential preprocessing steps, chief among them being normalization.
This critical process ensures consistent data quality and mitigates potential biases that may arise from variations in feature magnitudes.
We divide our dataset into three subsets: training (85\% of the dataset), validation (5\% of the dataset), and test (10\% of the dataset).

\subsection{Baseline Evaluation}
\label{subsec:baselineevaluation}

In this phase, we look at the evaluation of our fault prediction system.
Initially, we gauge the baseline performance of our system without incorporating countermeasures or exposure to adversarial attacks.
The training phase involves utilizing the training data, and subsequently, we evaluate the prowess of our \ac{gru}-based fault prediction system on the test set.
The results are notable, with our model achieving a mean accuracy of 0.604±0.01 for fault type prediction and an accuracy of 0.958±0.01 for fault zone prediction.
This performance marks a substantial improvement (a mean 33.11\% increase) compared to the state-of-the-art~\cite{mainpaper}.
When replicating this model from the literature, we implemented our preprocessing methods and adjusted the seed and computing environment to match those of our proposed model.
These modifications may have influenced the discrepancies observed between the reported results in the paper and our findings.
Also, we have chosen to integrate classical \ac{ml} algorithms such as XGBoost, Random Forest, and Decision Tree into our analysis for two primary reasons.
Firstly, they serve as a baseline for comparison against our proposed models, enabling us to gauge the performance and efficacy of our approaches. 
Secondly, their inclusion underscores the significance of considering causality between data points in this task, highlighting the importance of leveraging advanced techniques to capture temporal dependencies within the data.
The detailed results are presented comprehensively in Table~\ref{tab:accuracy}.

\begin{table}[h]
    \centering
     \caption{Comparison of the model's accuracy.}
    \begin{tabular}{l|c|c}
        \hline
        \multirow{2}{1cm}{\textbf{Model}} & \multicolumn{2}{c}{\textbf{Accuracy}} \\ \cline{2-3}
        & Fault Type & Fault Zone \\
        \hline
        MLP reproduced from~\cite{mainpaper} & 0.407 & 0.800 \\
        MLP claimed by~\cite{mainpaper} & 0.460 & 0.710 \\
        Decision Tree & 0.522 & 0.818 \\
        Random Forest & 0.543 & 0.831 \\
        XGBoost & 0.560 & 0.841 \\
        \textbf{GRU} & \textbf{0.604} & \textbf{0.958} \\
        \hline
    \end{tabular}
   
    \label{tab:accuracy}
\end{table}

\begin{figure}[!htpb]
  \centering
  \begin{subfigure}{0.495\textwidth}
     \centering
     \includegraphics[width=\textwidth]{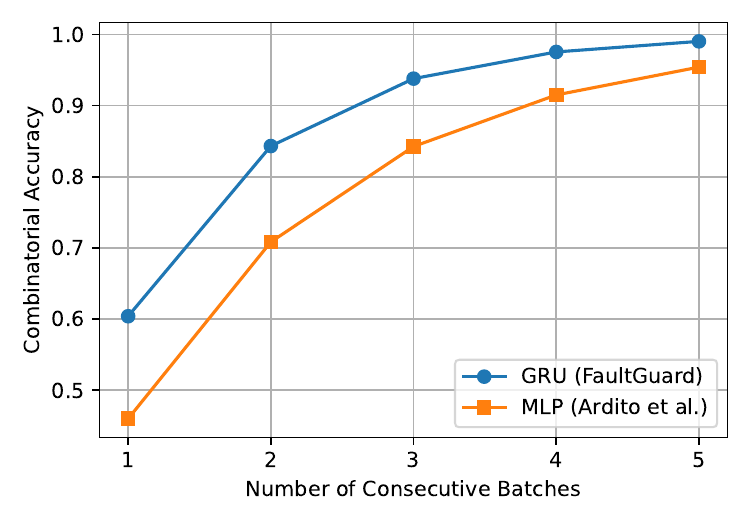}
     \caption{Fault type prediction.}
     \label{subfig:misclassification_type}
  \end{subfigure}
  \begin{subfigure}{0.495\textwidth}
     \centering
     \includegraphics[width=\textwidth]{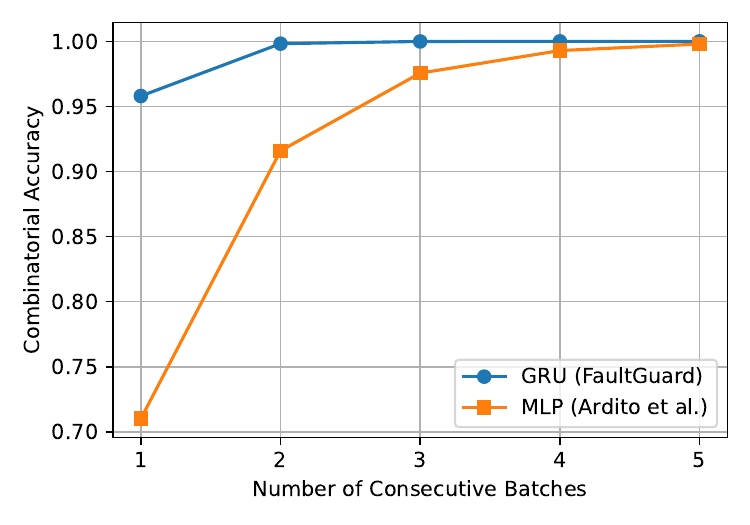}
     \caption{Fault zone prediction.}
     \label{subfig:misclassification_zone}
  \end{subfigure}
  \caption{Combinatorial accuracy of fault prediction tasks when delaying notification by the number of consecutive faulty batches.}
  \label{fig:misclassification}
\end{figure}

\paragraph{Combinatorial Accuracy.}
While our fault prediction system demonstrates superior evaluation accuracy compared to existing literature, the practical implementation necessitates a nuanced approach to address misclassification events.
Issuing notifications to grid operators for each detected fault could potentially result in multiple false alarms, leading to operational challenges.
To address this concern, we adopt a strategy where we wait for the identification of multiple consecutive fault data batches before triggering a notification.
This approach introduces the concept of \textit{combinatorial accuracy}, which considers the number of consecutive faulty batches required to initiate an alert.
This concept is crucial for balancing the trade-off between efficient fault detection and minimizing false alarms, ensuring system robustness in real-world scenarios.
The formula for combinatorial accuracy reflects a geometric distribution and is expressed as follows:
\begin{equation}
    combinatorial\_accuracy = \left( 1 - \left( 1 - accuracy \right)^{batches} \right).
\end{equation}
This formulation ensures that a higher accuracy value is associated with each notification, providing a more reliable indication of actual fault occurrences.
The relationship between this accuracy value and the number of consecutive faulty batches is thoroughly analyzed and illustrated in Fig.~\ref{fig:misclassification}, offering valuable insights into the system's performance under this combinatorial accuracy framework.
As evident, the simple strategy of awaiting confirmation from another faulty batch of data significantly enhances the model's accuracy for fault type prediction, improving from 0.604 to a score of 0.843.
Likewise, for fault zone classification, accuracy rises from 0.958 to 0.998.
This approach drastically reduces the probability of issuing a false alarm notification to grid operators.
Specifically, the probability is minimized to 13.7\% for fault type prediction and 0.2\% for fault zone prediction.
This underscores our methodology's effectiveness in elevating accuracy and mitigating the risk of generating false alarms, contributing to a more reliable fault prediction system.
Therefore, we opt for a notification delay parameter of two batches of unauthorized data, as it strikes a balanced trade-off between minimizing the false alarm rate and the timely data collection.

\subsection{Attacks Evaluation}
\label{subsec:attacksevaluation}

We now evaluate our attacks against the fault prediction models.
We divide the evaluation into the scenarios discussed in the threat model in Section~\ref{sec:systhreat}, i.e., white-box attacks and gray-box attacks.
\paragraph{White-box Evaluation.}
\label{subsec:WB}

\begin{figure}[!htpb]
  \centering
  \begin{subfigure}{0.495\textwidth}
     \centering
     \includegraphics[width=\textwidth]{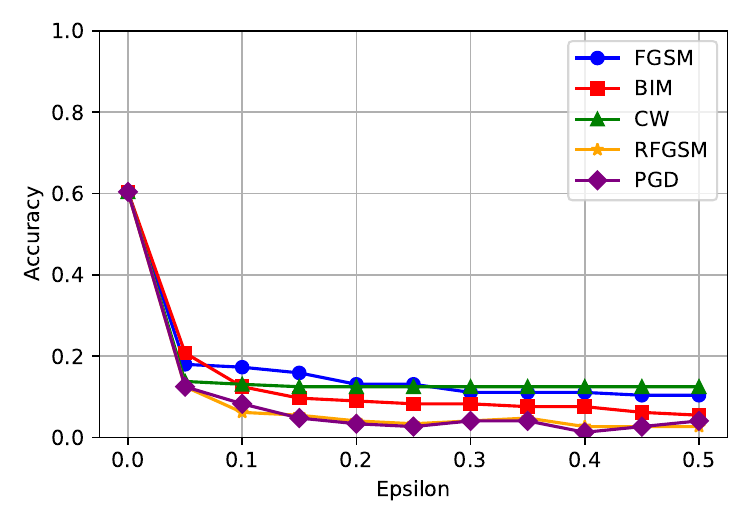}
     \caption{Fault type prediction accuracy.}
     \label{subfig:wbft}
  \end{subfigure}
  \begin{subfigure}{0.495\textwidth}
     \centering
     \includegraphics[width=\textwidth]{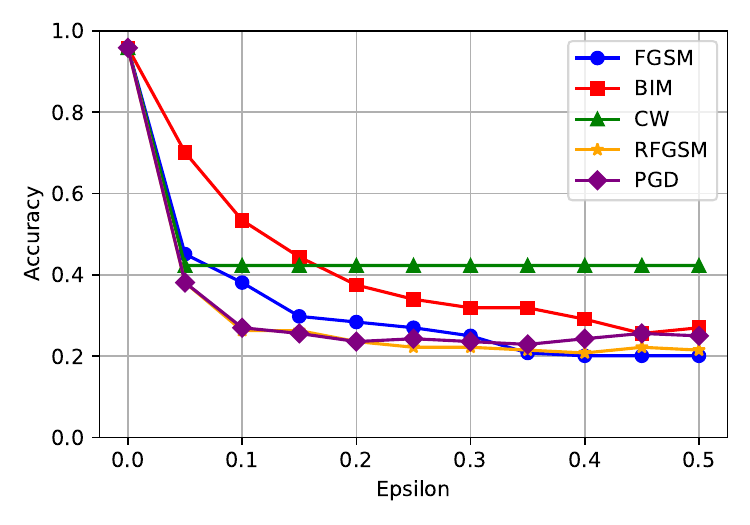}
     \caption{Fault zone prediction accuracy.}
     \label{subfig:wbfz}
  \end{subfigure}
  \caption{Model's accuracy at varying epsilon values on the white-box attacks.}
  \label{fig:wbaccuracy}
\end{figure}

In this evaluation, we comprehensively assess the efficacy of the white-box attacks, as discussed in Section~\ref{sec:attacks}.
To implement these attacks, we leverage the TorchAttacks library~\cite{kim2020torchattacks}, probing the baseline system to evaluate the susceptibility of our model without incorporating any countermeasures or defenses.
The attacks are executed with varying epsilon values, signifying the strength of each attack and the degree of perturbation introduced.
Specifically, we explore epsilon values ranging from 0.05 to 0.50.
The outcomes of these attacks across different tasks are visually presented in Fig.~\ref{fig:wbaccuracy}.
Notably, even with a minimal epsilon of 0.05, a significant decline in the performance of all models is evident.
In this case, the accuracy of the prediction model drops to (an average of) 0.155 for fault type and 0.467 for fault zone.
For the reproduced \ac{mlp} model from~\cite{mainpaper}, the accuracy drops to (an average of) 0.070 for fault type and 0.178 for fault zone under the FGSM attack.
This evaluation underscores the baseline system's vulnerability to white-box attacks, shedding light on the need for robust defenses and countermeasures to fortify our fault prediction model against adversarial threats.

\paragraph{Gray-box Evaluation.}
\label{subsec:GB}

In this section, we thoroughly examine the effectiveness of our gray-box attack, as detailed in Section~\ref{subsec:gray}.
This attack is executed using the generator component within our \ac{gan} model, originally designed for anomaly detection.
The key aspect of our approach is the use of binary cross-entropy loss, a fundamental element in the \ac{gan} framework.
The successful training of our \ac{gan} relies on specific hyper-parameters, including 150 training epochs and a learning rate of $2 \times 10^{-4}$, which governs the optimization process.
After completing the training of our \ac{gan}, we generate a considerable volume of synthetic malicious data, totaling around 1500 batches.
These synthetic data samples, crafted from random noise by our trained generator, are fed into our fully trained fault prediction system for classification.
The results show that the generator produces data that can be classified as belonging to 9 out of 11 classes in the fault type prediction task and 3 out of 4 classes in the fault zone prediction task.
Effectively, this translates to an \ac{asr} of 0.818 for fault type prediction and 0.750 for fault zone prediction.
This underscores that even without prior knowledge of the fault prediction models, an adversary with access to data can create deceptive, malicious data that can elude classic \ac{ads} employed in the smart grid. 

\subsection{FaultGuard Evaluation}
\label{subsec:faultguardevaluation}

In this section, we evaluate the performance of FaultGuard under the attacks described in the previous section.
First, we evaluate the effectiveness of the ADS module against white-box attacks. Then, we evaluate our ADS against white-box and gray-box attacks.

\paragraph{\ac{ads} Evaluation.}
In this section, we conduct an experimental evaluation of our \ac{ads}, a critical component of our defense strategy against \ac{gan}-based attacks and various white-box attacks, as detailed in Section~\ref{sec:delamain}.
We initiate the process by utilizing the training dataset to train our \ac{gan} model.
We retain the trained discriminator, a crucial element for our subsequent evaluation.
The evaluation involves merging generated malicious data with authentic test data from our dataset, simulating malicious attempts alongside real data.
These merged datasets are then subjected to the discriminator for analysis, utilizing a predefined threshold (0.5) for discerning determinations between legitimate and anomalous data.
The results underscore the discriminator's effectiveness, achieving a mean accuracy rate of $0.991 \pm 0.005$ standard deviation when classifying real data from malicious data generated by our \ac{gan} in the fault type prediction task and a mean accuracy rate of $0.972 \pm 0.005$ standard deviation in the fault zone prediction task.
We subject our model to different white-box adversarial attacks in the subsequent phase.
We generate adversarial attacks targeting the fault type and fault zone prediction models, individually applying these attacks to the models.
The results are detailed in Table~\ref{tab:evaluation1}, highlighting the \ac{ads}'s resilience and the contributions of our adversarial learning training layer.
The \ac{ads} achieves a mean accuracy rate of $0.979 \pm 0.050$ standard deviation when classifying real data from malicious data generated by our white-box attacks in the fault type prediction task and a mean accuracy rate of $0.821 \pm 0.050$ standard deviation in the fault zone prediction task.

\begin{table}[!htpb]
 \centering
    \caption{\ac{ads} accuracy for each attack and considering the Adversarial Learning (AL) layer. Results are averaged for each $\epsilon$ value.}
    \resizebox{\textwidth}{!}{
    \begin{tabular}{l|c|c|c|c|c|c}
        \hline
        \multirow{2}{*}{\textbf{Task}} & \multirow{2}{*}{\textbf{AL}} & \multicolumn{5}{c}{\textbf{Accuracy}} \\ \cline{3-7}
        & & FGSM & BIM & CW & RFGSM & PGD \\
        \hline
        \multirow{2}{*}{Fault Type} & \xmark & $0.674 \pm 0.032$ & $0.261 \pm 0.067$ & $0.386 \pm 0.006$ & $0.703 \pm 0.009$ & $0.700 \pm 0.007$ \\
         & \cmark & $\mathbf{1.000 \pm 0.000}$ & $\mathbf{1.000 \pm 0.000}$ & $\mathbf{0.897 \pm 0.000}$ & $\mathbf{1.000 \pm 0.000}$ & $\mathbf{1.000 \pm 0.000}$ \\
        \hline
        \multirow{2}{*}{Fault Zone} & \xmark & $0.274 \pm 0.032$ & $0.241 \pm 0.011$ & $0.232 \pm 0.000$ & $0.300 \pm 0.004$ & $0.293 \pm 0.008$ \\
        & \cmark & $\mathbf{1.000 \pm 0.000}$ & $\mathbf{0.999 \pm 0.001}$ & $0.108 \pm 0.002$ & $\mathbf{1.000 \pm 0.000}$ & $\mathbf{1.000 \pm 0.000}$ \\
        \hline
    \end{tabular}
    }
    \label{tab:evaluation1}
\end{table}

Despite the \ac{ads}'s robust performance against various white-box adversarial attacks, it exhibits vulnerability to the CW attack, a sophisticated adversarial technique known for its intricacy.
While the \ac{ads} may occasionally miss some batches of adversarial attacks, particularly those crafted using the intricate CW technique, the subsequent layer of defense comes into play.
The adversarial learning layer ensures the fault prediction models' resilience against these advanced attacks.
In only one case (i.e., fault zone + AL against CW attacks), we notice a drop in performance with the inclusion of our novel layer.
This standalone case is caused by the missing inclusion of the CW attack in the layer, and more details are given in Section~\ref{sec:takeaways}.
As elucidated in upcoming sections, our fault prediction models showcase remarkable resilience even when confronted with the CW attack, successfully averting a significant decline in system performance.
The multi-layered defense strategy underscores a comprehensive approach aimed at enhancing the overall robustness and effectiveness of the fault prediction system in smart electrical grids.

\begin{figure}[!htpb]
  \centering
  \begin{subfigure}{0.495\textwidth}
     \centering
     \includegraphics[width=\textwidth]{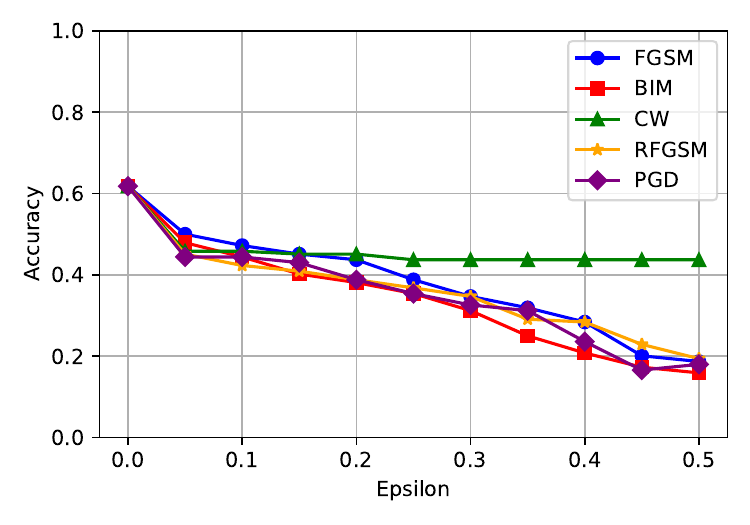}
     \caption{Fault type prediction accuracy.}
     \label{subfig:wbftd}
  \end{subfigure}
  \begin{subfigure}{0.495\textwidth}
     \centering
     \includegraphics[width=\textwidth]{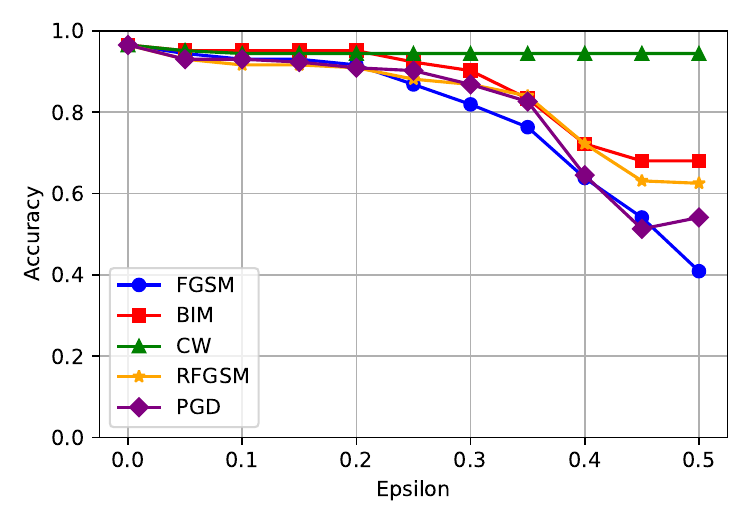}
     \caption{Fault zone prediction accuracy.}
     \label{subfig:wbfzd}
  \end{subfigure}
  \caption{Model's accuracy at varying epsilon values on the white-box attacks when equipped with the defense mechanism.}
  \label{fig:wbaccuracyd}
\end{figure}

\paragraph{Fault Prediction Model Evaluation.}
This section comprehensively evaluates our fault prediction systems, augmented with online adversarial training as a defense mechanism.
As observed in Section~\ref{subsec:WB}, our model, while surpassing the state-of-the-art, remains vulnerable to white-box adversarial attacks.
We introduce a novel training layer to fortify our models against such attacks, extensively discussed in Section~\ref{subsec:Faultgurad}.
After integrating this new training layer, we assess our models' performance in fault type and zone prediction critical tasks.
The outcomes of this evaluation are presented in Fig.~\ref{fig:wbaccuracyd}.
Notably, the resilience of our models against adversarial attacks, particularly in comparison to models without the defense mechanism, exhibits a substantial improvement.
This enhancement is particularly pronounced when facing sophisticated attacks like CW, which have demonstrated the ability to bypass our ADS system. 
The results underscore the efficacy of the introduced adversarial training layer in bolstering the model's robustness, significantly mitigating the impact of adversarial attacks on fault prediction performance.
This defense mechanism is a crucial safeguard, ensuring our fault prediction systems' continued reliability and effectiveness, even in the face of sophisticated and intricate adversarial challenges.
A summary of the results of our models and their improvement when paired with the defense mechanism is shown in Table~\ref{tab:evaluation}.

\begin{table}[h]
    \centering
    \caption{Comparison of model performance under adversarial attacks ($\epsilon=0.05$) before and after employing our Online Adversarial Training (OAT).}
    \begin{tabular}{l|c|c|c|c|c|c|c}
        \hline
        \multirow{2}{*}{\textbf{Task}} & \multirow{2}{*}{\textbf{OAT}} & \multicolumn{6}{c}{\textbf{Accuracy}} \\ \cline{3-8}
        & & Baseline & FGSM & BIM & CW & RFGSM & PGD \\
        \hline
        \multirow{2}{*}{Fault Type} & \xmark & 0.604 & 0.180 & 0.208 & 0.138 & 0.125 & 0.125 \\
        & \cmark & \textbf{0.618} & \textbf{0.500} & \textbf{0.479} & \textbf{0.458} & \textbf{0.451} & \textbf{0.444} \\ \hline
        \multirow{2}{*}{Fault Zone} & \xmark & 0.958 & 0.451 & 0.701 & 0.423 & 0.381 & 0.381 \\
        & \cmark & \textbf{0.965} & \textbf{0.944} & \textbf{0.951} & \textbf{0.951} & \textbf{0.930} & \textbf{0.930} \\
        \hline
    \end{tabular} 
    \label{tab:evaluation}
\end{table}

\paragraph{Computational Cost.}
All experiments in this paper have been conducted on Kaggle as a cloud resource with the following configurations: Intel Xeon (2.20 GHz), NVIDIA Tesla P100 (3584 Cuda cores, 16 GB), 30 GB of RAM, Linux Debian with Python 3.10.12\footnote{Additional details on packages' versions are available at: \url{https://anonymous.4open.science/r/FaultGuard-1518/requirements.txt}}.
The defense strategies have shown commendable performance through lightweight models and two white-box adversarial attacks.
However, a notable trade-off is the introduction of additional computational load.
The \ac{ads} and fault prediction systems have been augmented with two adversarial data generators and incorporated two new loss calculations for the prediction models to enhance robustness.
As a result, there is a significant increase in computational demands, leading to longer training times for the models.
The changes significantly affect training times, as depicted in Table~\ref{tab:training_times}, comparing standard models to those fortified with defense mechanisms.
Despite the increased computational cost, it is crucial to consider the enhanced robustness and resilience brought by these defenses.
Furthermore, this training procedure is needed only when initially deploying the model in the system and does not require maintenance. 
This computational trade-off underscores the ongoing challenge of balancing model efficiency with the imperative to defend against adversarial threats in the smart grid domain.
\begin{table}[h]
 \centering
    \caption{Model training times with and without Online Adversarial Training (OAT).}
    \begin{tabular}{l|c|r}
        \hline
        \textbf{Task} & \textbf{OAT} & \textbf{Training Time} \\
        \hline
        \multirow{2}{*}{Fault Type} & \xmark & 1.18 min \\
        & \cmark & 78.90 min \\ \hline
        \multirow{2}{*}{Fault Zone} & \xmark & 1.29 min \\
        & \cmark & 95.00 min \\
        \hline
    \end{tabular}
    \label{tab:training_times}
\end{table}
\section{Takeaways}
\label{sec:takeaways}

Fault prediction systems are a well-researched topic in the literature.
However, researchers often neglect the security aspect of these systems, making their implementation problematic due to the high chances of errors when dealing with adversaries.
Therefore, we present a summary of the key takeaway messages, making it easier for practical implementation in real-world scenarios.
By doing so, we assist practitioners in effectively applying these systems and offer researchers recommended best practices for their studies.

\begin{formal}
\textbf{Takeaway 1} -- \textit{Fault prediction systems are vulnerable to adversarial attacks, regardless of the scenario's assumptions on the attacker's knowledge.}
\end{formal}

As discussed in Section~\ref{subsec:attacksevaluation}, adversarial attacks are particularly effective against the models tested in this study.
Indeed, model accuracy dropped by up to 74.34\% with an epsilon value of just 0.05.
While these values are valid only for white-box scenarios, it is worth noting that even in gray-box scenarios (i.e., having access to the data), the attacker has great leverage on the system, reaching \acp{asr} up to 0.818.
As such, adversarial attacks require particular consideration when designing a \ac{ml}-based fault prediction system.

\begin{formal}
\textbf{Takeaway 2} -- \textit{Complex attacks, such as CW, are more difficult to be detected by the \ac{ads}.}
\end{formal}

As shown in Table~\ref{tab:evaluation1}, our \ac{ads} can detect most attacks with perfect accuracy, except for the CW attack.
While in the fault type prediction task we obtained an accuracy of 0.897 with the addition of the adversarial learning layer, in the fault zone task, including this layer effectively worsened its performance (from 0.232 to 0.108).
It is noteworthy that adversarial learning is performed only with the FGSM and BIM attacks.
However, even when including CW in the training process, there were no improvements w.r.t. the standard \ac{ads} model against the same attack.
For these reasons, complex attacks require special attention, as improperly implementing the \ac{ads} might not be enough to detect them.

\begin{formal}
\textbf{Takeaway 3} -- \textit{Complex attacks, such as CW, are more effective against good-performing models.}
\end{formal}

Another correlation regarding the CW attack that can be extracted from our evaluation is that attacks are more effective against good-performing models.
Indeed, the fault type prediction model (baseline accuracy of 0.604) was less affected by the attacks w.r.t. the fault zone prediction model (baseline accuracy of 0.958).
As such, while researchers and practitioners strive to reach the highest possible accuracy score, the threat of adversarial attacks becomes more pronounced.
This highlights the paradoxical relationship between model performance and susceptibility to attacks.
Consequently, the pursuit of high accuracy should be accompanied by a heightened awareness of potential vulnerabilities and the implementation of robust defense mechanisms.

\begin{formal}
\textbf{Takeaway 4} -- \textit{Different forms of adversarial training greatly improve the models' resistance against adversaries.}
\end{formal}

One of the significant contributions of this paper is the proposal of novel adversarial training techniques.
Indeed, including the adversarial learning layer on the \ac{ads} significantly improved its performance, and the online adversarial training performed on the fault prediction model made it more resilient against attacks.
These findings underscore the effectiveness and versatility of diverse adversarial training methodologies in improving model robustness and defense capabilities, contributing valuable insights to advancing secure and resilient machine learning models.
\section{Conclusions}
\label{sec:conclusions}

In the smart grid, fault prediction systems are promising tools that can ensure energy delivery and reliability.
However, despite the growing interest in the literature, the security aspect of these systems is often neglected.
This oversight can lead to safety issues and delays, making the implementation of those systems counterproductive.
\paragraph{Contribution.}
In this paper, we introduced \textbf{FaultGuard}, a resilient framework designed for fault type and zone classification tasks.
To ensure the security of our system, we incorporated an \ac{ads} with a unique \ac{gan} training layer to detect attacks.
Additionally, our approach involved a low-complexity fault prediction model and employed an online adversarial training technique to bolster robustness.
We thoroughly evaluated the framework's performance, assessing its resilience against diverse adversarial attacks using the publicly available IEEE13-AdvAttack dataset, a simulated dataset derived from the IEEE-13 test node feeder.
FaultGuard outclassed the state-of-the-art, reaching accuracy values up to 0.958 and being natively resilient against adversarial attacks.
Furthermore, our \ac{ads} detected attack attempts with accuracies up to 1.000.
\paragraph{Future Work.}
While still outclassing the state-of-the-art and other \ac{ml} models in the same task, improving the accuracy for fault type prediction is needed to further strengthen its contribution.
Indeed, the accuracy value achieved in this study is highly dependent on the dataset employed.
As such, creating a richer dataset is the most significant way to improve these results.
Also, since this paper focused on evasion attacks towards fault prediction systems, studying poisoning attacks might provide powerful insights into the resilience of these systems.
By delineating a new system and threat model accounting for this threat, it might be possible to study the vulnerabilities of the models and thus improve their security.
Finally, by combining these results with those obtained in this paper, we would be able to provide a complete overview of the resilience of \ac{ml}-based fault prediction models, aiding practitioners in safely deploying these systems.

\bibliographystyle{splncs04}

\bibliography{bibliography}

\end{document}